\documentclass[runningheads]{llncs}
\usepackage{uppaal}
\usepackage{amsmath,amssymb,dsfont,stmaryrd}
\usepackage{mathpartir}
\usepackage{tikz}
\usepackage{ifthen}
\usepackage{multirow}
\usepackage{algpseudocode}
\usepackage{enumitem}
\usepackage{xstring}
\usepackage{tikz-cd}
\usepackage{subcaption}
\usepackage{algorithm}
\usepackage{centernot}
\usepackage[hidelinks,bookmarks,bookmarksopen,bookmarksdepth=3]{hyperref}
\usepackage[utf8x]{inputenc} 
\usepackage{environ}

\input{main.sty}

\title{Energy Consumption Optimization in Radio Access Networks
  (ECO-RAN)\thanks{In 2022 ECO-RAN wins Project of the Year at Energy
    Cluster Denmark.}
  \\
  \small{Technical Report} }

\author{%
  Anders~Mariegaard
  \and
  Kim~G.~Larsen
  \and
  Marco~Mu\~{n}iz
  \and
  Thomas~Dyhre~Nielsen
  \institute{%
    Department of Computer Science, Aalborg University, Denmark
  \\
  \email{\{am,kgl,muniz,tdn\}@cs.aau.dk}}}

\authorrunning{Mu\~{n}iz et al.}
\titlerunning{ECO-RAN}
\begin{document}
\maketitle
\begin{abstract}



  In recent years, mobile network operators are showing interest in
  reducing energy consumption.  Toward this goal, in cooperation with
  the Danish company 2Operate we have developed a stochastic
  simulation environment for mobile networks. Our simulator interacts
  with historical data from 2Operate and allow us to turn on and off
  network cells, replay traffic loads, etc.
  We have developed an optimization tool which is based on stochastic
  and distributed controllers computed by \uppaal. We have conducted
  experiments in our simulation tool. Experiments show that there is a
  potential to save up to 10\% of energy. We observe that for larger
  networks, there exists a larger potential for saving energy.
  Our simulator and \uppaal controllers, have been constructed in
  accordance to the 2Operate data and infrastructure. However, a main
  difference is that current equipment do not support updating
  schedulers on hourly bases. Nevertheless, new equipment e.g.\ new
  Huawei equipment do support changing schedulers on hourly
  basis. Therefore, integrating our solution in the production server
  of 2Operate is possible. However, rigorous testing in the production
  system is required.
  %
  
\end{abstract}

\section{Introduction}

In accordance with the enormous expansion of mobile networks in
Denmark and the rest of the world, the number of mobile masts
providing coverage has exploded, and with the upcoming expansion of
5G, there will be even more mobile radio devices that require power.

In recent years, it has been in the interest of the mobile operators
to bring the power consumption, and the first steps have already been
taken.  These measures are based on semi-automatic procedures and with
strong assumptions e.g.\ everyone follows the same patterns.
A more fully automated approach to the problem, based on artificial
intelligence, is desirable and expected to be able to further reduce
power consumption.
Furthermore, in connection with the sales activities, both in and
outside Europe, it has been made clear that the mobile operators are
increasingly concerned about mobile network power consumption, now and
especially in the future.  The background for this is that electricity
consumption in the mobile network will increase significantly with the
introduction of 5G and several 4G frequency layers.  There is
therefore both a considerable financial gain by minimizing power
consumption, and also a growing interest in contributing to the Danish
climate action, where the goal is for Denmark to be energy-renewable
in 2050.

There are thus already good market leadership advantages for companies
that can demonstrate that they are actively making an effort to
achieve this goal.  Together with one of our partners 2Operate we
carried out feasibility studies that show that some Nokia-specific
functions in Nokia's operating system can force the radio units to
switch off at certain times -- e.g.\ at 01:00-06:00. That is it is
possible to synthesize and implement schedulers which turn on and off
power cells.

Conservative calculations show that the savings potential will be €100
– €300 annually per mobile mast per company. In Denmark, they have
fire mobile operators together approx. 10,000 locations with their
equipment.  This gives a total savings potential of DKK 1 - 3
million. euros annually or up to 50 million KWh. Worldwide, this
potential will be many times higher.


\section{Preliminaries}

\subsection{Mobile Networks}

\begin{figure}[t]
    \centering
    \includegraphics[width=\textwidth]{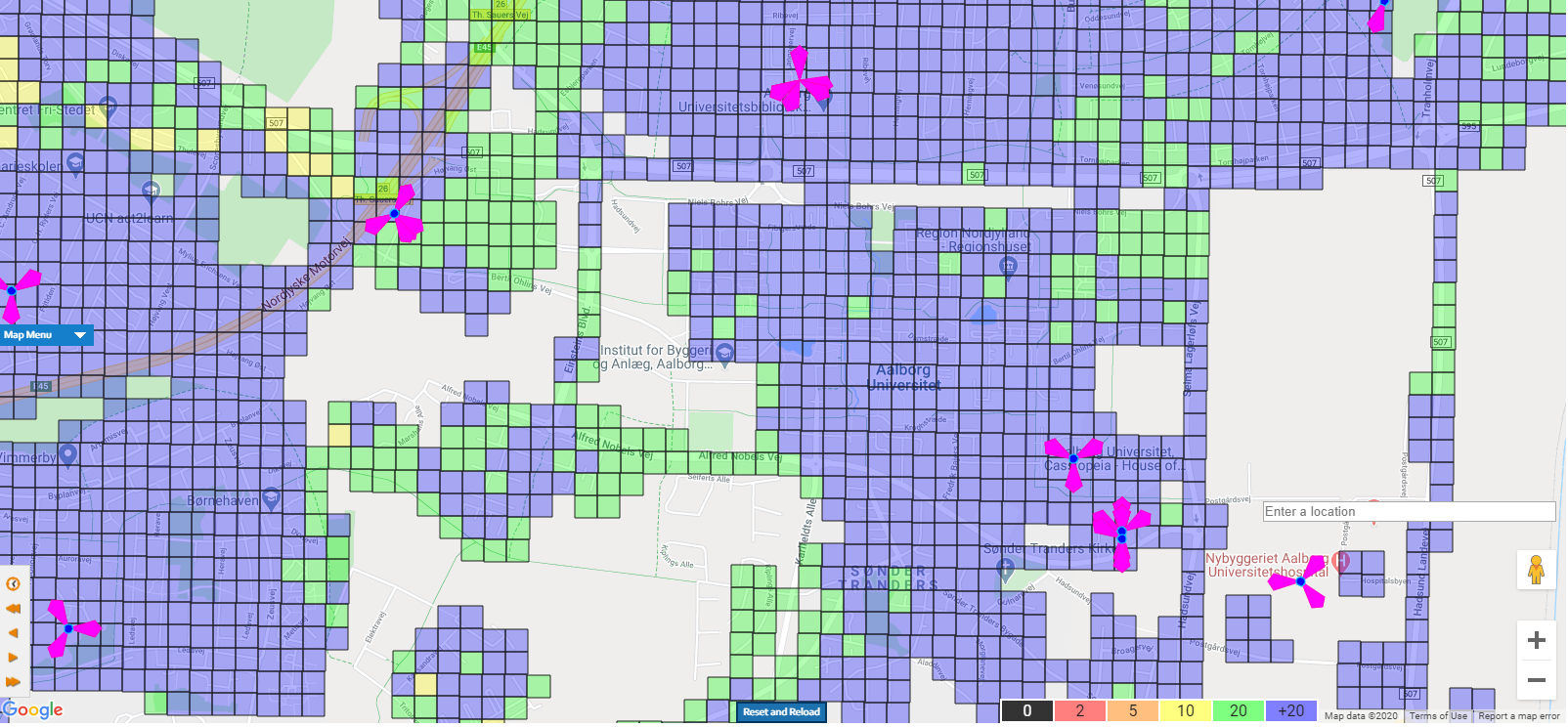}
    \caption{Base station with three sectors. Sector 1 consists of one 800 cell and one 1800 cell.}
    \label{fig:networkmap}
  \end{figure}

In this work we consider some geographical area where there is a
number of \emph{base stations}. Base stations have number of
\emph{cells} and every \emph{cell} operate in some \emph{frequencies}.
The geographical location is discredited using \emph{pixels}. Cells
provide coverage to pixels and each pixel has a traffic demand.
Figure~\ref{fig:networkmap} shows a map with the building of the
Department of Computer Science at Aalborg University. Base stations
are in pink, every base station contain some cells, and pixels
correspond to the grid elements.

\subsubsection{Frequency Layers}
Each base station usually consists of a number of cells broadcasting
on different frequencies. Lower frequencies are for coverage while
higher frequencies are for capacity. For 4G, the 800 MHz frequency
layer is considered the coverage layer and must not be turned off in
the current setup. There is room for optimization at the higher
frequency layers as the needed capacity fluctuates a lot during a
typical day.  The 4G (LTE) frequencies are:
  E: 800 MHz,
  V: 900 MHz,
  T: 1800 MHz,
  A: 2100 MHz,
  L: 2600 MHz. 

\subsubsection{Power Saving}

The goal is to shut down capacity layers then they are not
needed. e.g.\ during the night. A current contraint of the system is
to maintain coverage to all pixels. In the current system, to ensure this constraint the 800
MHz layer can not be shut down.


\subsubsection{Historical Data}

The company 2Solve has relevant historical data, e.g.\ the traffic
demands for every base station. There is also information about the
signal strengh for every pixel and for every frequency layer. Our
simulations will be based on existing historical data.


\section{Computing Near Optimal Strategies}

\tikzstyle{tool}=[draw, fill=blue!30, text width=3cm, 
text centered, minimum height=2.5cm]
\tikzset{
  empty/.style={rectangle,thick,draw=none,fill=none},
}

\subsection{Optimal Controlling}

The main mathematical formalism for our modeling and posterior
optimization is a stochastic hybrid game. For details we refer the
reader to~\cite{DBLP:conf/tacas/LarsenMMST16}.
At a high level the game is between a controller and the environment.
In our concrete scenario the stochastic game $\G$ corresponds to a
communication network in which the environment consists of a number of
cells $\nrCells$, pixels $\nrPixels$ and traffic demand per pixel. The
controller consists of modes \textsf{ON} or \textsf{OFF} for every
cell. Given a time horizon $H$ e.g.\ of one day, a \emph{control
  strategy} $\sigma^H$ for horizon $H$, determines if a given cell is
\textsf{ON} or \textsf{OFF}. The stochastic of the system come from
the traffic demands, which can be represented as probability
distributions over the pixels. Note that given a stochastic hybrid
game $\G$ and control strategy $\sigma^H$, the game under the
strategy $\G_{\nrCells} \upharpoonright\sigma^H$ is a stochastic
process which induces a measure on the possible executions of the
system.

\begin{definition}[Optimal Controlling]
  \label{def:optimal}
  Given a stochastic hybrid game $\G_{\nrCells}$, synthesize a
  strategy $\sigma^H$ which minimizes the expected reward 
  \[
     \sigma^H = 
     \mbox{{\sf argmin}}_{\sigma}\E^{\G_{\nrCells}}_{\sigma,H}(\reward) 
   \]
   where $\reward$ accumulates the energy usage and a penalty for the
   lack of coverage 
    \[
    \reward = \int_0^H
    \penal(t) + \energy(t) \ dt
  \]
  with 
  \begin{align*}
      \penal(t) & = \sum^{\nrPixels}_i \penal(t,i)
      \\
      \penal(t,i)& =
      \begin{cases}
        0 & \text{ if } \contribution(t,i) - \demand(t,i) > 0 
      \\
        1000 & \text{ otherwise }
      \end{cases}
  \end{align*}
  and
  
    \begin{align*}
      \energy(t) &= \sum^{\nrCells}_i \energy(t,i) \\
      \energy(t,i) &=
                     \begin{cases}
                       0 & \text{ if cell is off}
                       \\
                       \text{cell power + cost per mb}
                     \end{cases}
    \end{align*}
    $\contribution(t,i)$ indicates the cells contribution to pixel $i$
    at time $t$ similarly $\demand(t,i)$ indicates the demand for
    pixel $i$ at time $t$.
\end{definition}

In this project we aim to control real world communication networks
with hundreds of cells and millions of pixels. Therefore computing the
strategy $\sigma^H$ is intractable. Instead we will compute near
optimal-strategies using diverse techniques.

\subsection{Online Strategy Synthesis}%
\label{onlineStrategySynthesis}
For this case study our goal is to compute a strategy (controller)
$\sigma^H$ to minimize energy consumption for a long horizon $H$. As
the number of choices for the controller grows exponentially in the
horizon, computing the strategy for a long horizon $H$ is
intractable. To overcome this problem we resort to the \emph{Online
  Strategy Synthesis}~\cite{DBLP:conf/tacas/LarsenMMST16} methodology,
where we periodically compute a online strategy $\sigma^h$ for a short
horizon $h < H$. By composing the online strategies $\sigma^h$ we can
control the system for the horizon $H$.  The composed strategy is
less optimal than the optimal strategy $\sigma^H$ but it can be
computed effectively.

Figure~\ref{fig:onlinestrategysynthesis} shows the online strategy
synthesis approach for $n$ \emph{cells}, a horizon $H$ of 1 day and
controlling every 60 min. Short horizon $h$ of 180 minutes.  For $n$
\emph{cells} for the offline controllers $\sigma^H$ and online
controller $\sigma^h$ there are \ $2^{720n}$ vs.\ $2^{3n}$
decisions. Thus computing a near-optimal online controller $\sigma^h$
is clearly more applicable.

The methodology has successfully been applied to multiple
case studies involving cyber-physical systems such as, intelligent
traffic lights~\cite{StrategoItsEurope17}, floor heating
systems~\cite{DBLP:conf/tacas/LarsenMMST16},
rerouting~\cite{doi:10.1177/03611981211000348} etc.

 \newcommand{\useStrategy}[4]{%
   \draw[thick, dotted] (#1,#2) -- (#3,#4);
 }

\begin{figure}[t]
  \centering
  \scalebox{0.7}{
    \begin{tikzpicture}[x=1.5cm]

        \draw [thick] (-5, -2.5) -- (1.2, -2.5);
        \draw [thick] (2, -2.5) -- (4, -2.5)
        node[above right] {$\scriptstyle minutes$};
        \node [empty] at (1.65,-2.5) {$\cdots$};

        \foreach \x/\l in {-5/0,-4/60,-3/120,-2/180,-1/240,0/300,1/360,3/{$H-60$},4/$H$} {
          \draw (\x, -2.4) -- (\x, -2.6);
          \node [empty] at (\x,-2.8) {\small \l};
        }

      
      \draw [thick,->] (-5, -1.5) -- (4, -1.5)  node[above right]
      {$\sigma^H$};

      \draw [thick,dashed,-] (-5, -0.5) -- (-4, -0.5); 
      \draw [thick,->] (-4, -0.5) -- (-1, -0.5)  node[above
      right] {$\sigma^h_0$}; 
      \draw [thick,dashed,-] (-4, 0.5) -- (-3, 0.5);
      \draw [thick,->] (-3, 0.5) -- (0, 0.5)  node[above
      right] {$\sigma^h_1$}; 
      \useStrategy{-4}{-0.5}{-4}{-2.5};
      \draw [thick,dashed,-] (-3, 1.5) -- (-2, 1.5);
      \draw [thick,->] (-2, 1.5) -- (1, 1.5)  node[above right] {$\sigma^h_2$};
      \useStrategy{-3}{0.5}{-3}{-2.5};
      \useStrategy{-2}{1.5}{-2}{-2.5};
    \end{tikzpicture}
  }
  \caption{Online Strategy Synthesis for $n$ \emph{cells}, a horizon
    $H$ of 1 day and controlling every 60 min. Short horizon $h$ of
    180 minutes.  For $n$ \emph{cells} for the offline controllers
    $\sigma^H$ and online controller $\sigma^h$ there are $2^{3n}$
    vs.\ $2^{720n}$ decisions.  }
  \label{fig:onlinestrategysynthesis}
\end{figure}
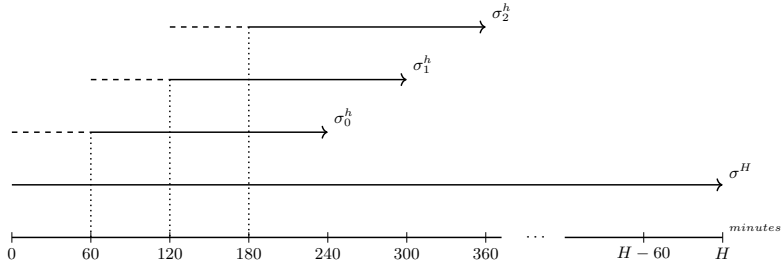

\begin{figure}[b]
  \centering
  \scalebox{0.8}{
    \begin{tikzpicture}
      \node (uppaal) at (10,0) [tool] {\stratego{}};
      \node (world) at (0,0) [tool] {Real World \\ (simulator)};

      \node (worlddes) at (1,-3) [empty] {{\parbox{8.0cm}{
            \flushleft
            \begin{itemize}
            \item
              Stochastic Game $\G$
            \item
              Control strategy $\sigma^H$ for horizon $H$
            \item
              Number of cells $\nrCells$
            \item
              Number of pixels $\nrPixels$
            \item
              {Python implementation}
            \item
              Connects to 2Operate data base
            \end{itemize}
          }}};

      \node (uppaaldes) at (11,-3) [empty] {{\parbox{8.0cm}{
            \flushleft
            \begin{itemize}
            \item
              Uppaal Model $\mathcal{G}'$
            \item
              Control strategy $\sigma^h$ for {\bf short} horizon $h$
            \item
              Number of cells $\nrCells$
            \item
              Number of pixels $\nrPixels$
            \item
              \stratego{} with C libraries
            \end{itemize}
          }}};

      \path[->,thick,black,draw,transform canvas={yshift=2mm}]
      (world) 
      edge node [above,pos=0.5] {\parbox{4.0cm}{\flushleft
          Status of the world \\ forecast traffic demand}}         
      (uppaal);
      \path[->,thick,black,draw,transform canvas={yshift=-2mm}]
      (uppaal) 
      edge node [below,pos=0.5,yshift=4mm, xshift=1mm] 
      {\parbox{4.0cm}{\flushleft
          control strategy $\sigma^h$}}         
      (world);
    \end{tikzpicture}
  }

  \caption{System Architecture}
  \label{fig:system}
\end{figure}

\subsection{Distributed Online Synthesis}

In this project we aim to control large scenarios with hundreds of
cells and millions of pixels. Therefore, directly applying online
strategy synthesis is not scalable. To overcome this difficulty, we
apply Distributed Online Synthesis as
in~\cite{DBLP:conf/tacas/LarsenMMST16}. Given a geographical area with
hundreds of cells, we partition it to sub areas which contain at most 8
cells. Then we can compute a online strategy for every partition and
then merge the strategies to control the full network.

\subsection{Methodology}

The real world consists of a number of base stations, cells, pixels,
frequency layers, etc.\ Where the goal is to provide a
\emph{controller} that powers ON or OFF cells to save energy. There
exist a number of tools which can be used to simulate the behavior of
mobile networks.

Figure~\ref{fig:system} illustrates our methodology. The real world
(or a simulation) require a control or a \emph{strategy} $\sigma^H$
for minimizing energy consumption for a long horizon $H$ e.g.\ 3
months. Since the number of choices for the controller grow
exponentially on the horizon $H$, computing a ``global'' strategy
$\sigma^H$ is not possible. Instead we periodically monitor the system
and compute a near-optimal strategy $\sigma^h$ for a short horizon
e.g.\ 3 hours. In this work we will use \stratego{} to compute online
strategies $\sigma^h$.


\section{Experimental Evaluation}

\subsection{Simulation Tool}
\begin{figure}[t]
  \centering
  \includegraphics[scale=0.7]{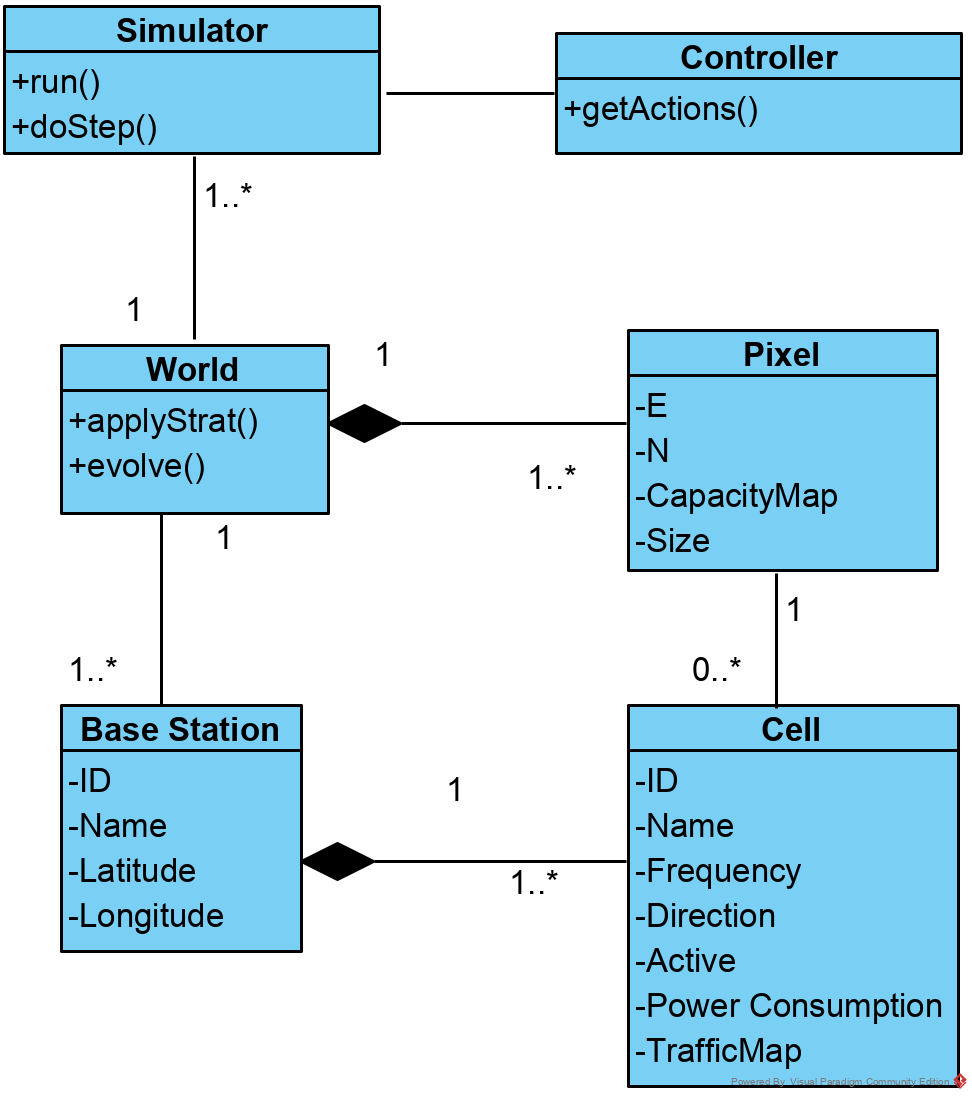}
  \caption{Simulator architecture}
  \label{fig:simulatorclasses}
\end{figure}

We use a simulation tool written in Python to replay and simulate
historical data. From historical data we can observe the coverage
contributions of every cell to every pixel. Then we can use this
information to reproduce the effects of turning \textsf{ON} or
\textsf{OFF} a given cell. In this way we can compute the values
required by Definition~\ref{def:optimal}. The traffic demand is based
on historical data with the additional assumption that the demand is
uniformly distributed over all pixels affected by
cell. Figure~\ref{fig:simulatorclasses} shows the overall architecture
of the network simulator.

\subsection{\stratego Controller}
\label{def:strategocontroller}

The main contribution of our work is to synthesize a near-optimal
strategy in accordance to Definition~\ref{def:optimal}. Toward this
goal we use the tool \stratego{}~\cite{uppaal}. The tool developed at
Aalborg University and used to facilitate generation and optimization
of strategies for abstract games with stochastic and real-time
aspects. The tool uses simulation-based statistical machine learning
methods.

\begin{figure}[t]
  \centering
  \includegraphics[scale=0.7]{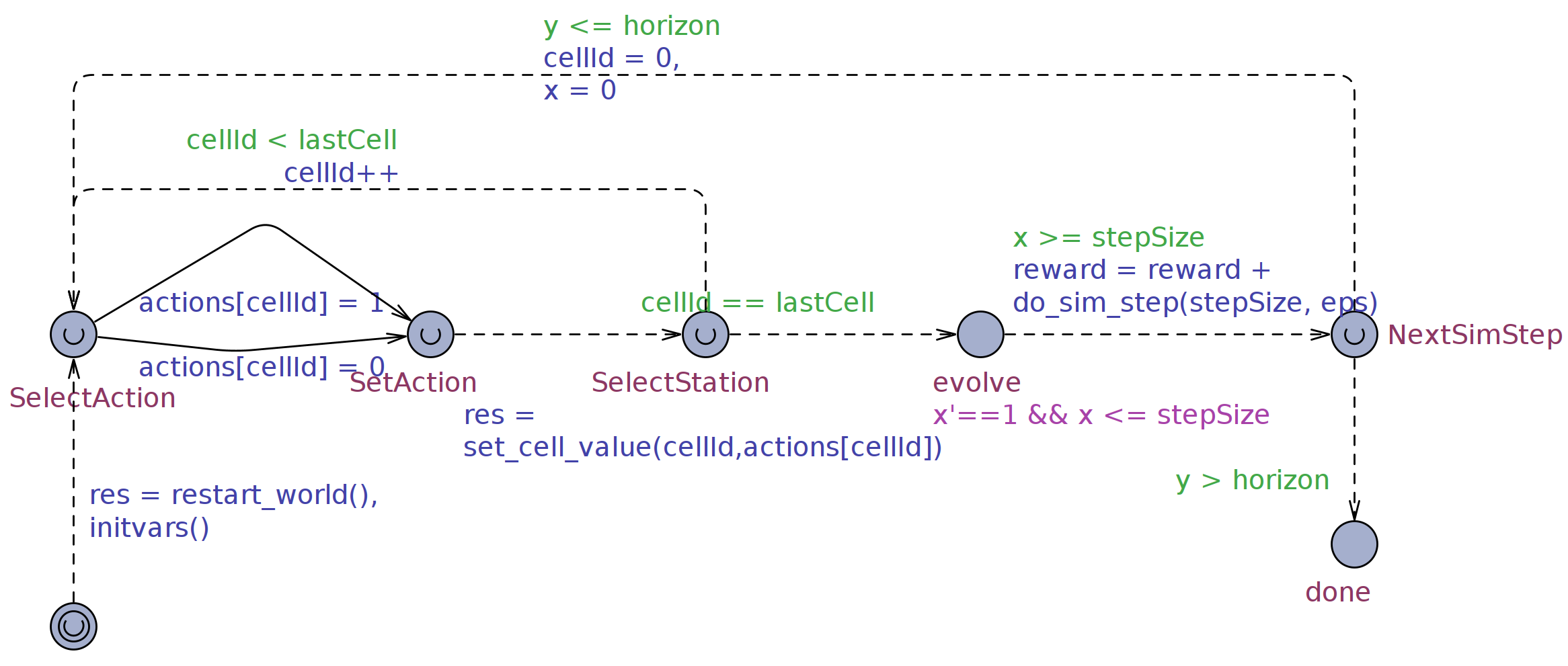}
  \caption{\stratego Controller}
  \label{fig:strategocontroller}
\end{figure}

Figure~\ref{fig:strategocontroller} shows the \stratego model for a
stochastic hybrid game $\G_{\nrCells}$. Solid arrows correspond to
controllable actions where as dashed arrows correspond to environment
actions. At every simulation step and for every cell the controller
has the choice to set \textsf{ON} or \textsf{OFF} the given cell,
indicated by the command \uppUpdate{actions[cellId]=1} or
\uppUpdate{actions[cellId]=0}. Once actions on cells have been chosen
the environment executes its actions, this is done by calling a
external C library with the command
\uppUpdate{do\_sim\_step(stepSize,eps)}.  This function return a real
value which is then accumulated in the variable \uppVar{reward}. These
steps are then executed until the short time horizon $h$ has been
reached. \stratego will perform a number of simulations and used
machine learning techniques to find the controllable actions which
optimize the variable \uppVar{reward}. Once the learning is complete
\stratego returns the near-optimal strategy which is then implemented
in the simulator (or the real world).

\subsection{Simulation Scenarios and Controllers}

As a proof of concept we have chosen to perform a simulation of 1 day
in the following two geographical locations in Aalborg, Denmark:

\begin{itemize}
\item City Syd with 39 cells and 2687 pixels   
\item Frydendal - Nørre Tranders  with 107 cells and 6138 pixels
\end{itemize}  

In our experiments we have used the following controllers:

\begin{itemize}
\item \emph{ALLON} all cells always ON
\item \stratego as described in Definition~\ref{def:strategocontroller}
\end{itemize}

Table~\ref{tab:results} shows the results of the different controllers
in the different scenarios. The columns energy, penalty and reward
correspond to Definition~\ref{def:optimal}. The values on column
energy are computing using a linear function on historical data and a
constant for a cost per megabit. 

We observe that while having no penalty, the \stratego controller is
able to save about 10\% energy on a single day. Concerning controller
\stratego, the computation time for strategies about 22 and 31 hours
for City Syd and Frydendal - Nørre Tranders on 16 cores. This means
that given sufficient hardware resources, the scenarios could be
controlled in real time. This is because using Online Strategy
Synthesis (c.f.\ Section~\ref{onlineStrategySynthesis}) will give a
time window of up to 60 minutes to compute the next near optimal
strategy.

\begin{table}[t]
  \centering
  \begin{tabular}{|l||c|c|c||c|c|c|}
    \hline
    \multirow{2}{*}{Scenario}  &  \multicolumn{3}{c||}{\emph{ALLON}}
    &  \multicolumn{3}{c|}{\stratego\footnote{}}
    \\\cline{2-7}
                               & Energy & Penalty & Reward M. & Energy & Penalty & Reward M.
    \\ \hline
    City Syd
                               & 3473&0&150  & 3191&0&141 
    \\ \hline\hline
    \parbox{3cm}{Frydendal \\ Nørre Tranders}
                               & 10115 & 0 & 436   & 9347&0&394

    \\ \hline
  \end{tabular}    
  
  \caption{Experimental results}
  \label{tab:results}
\end{table}


\section{Conclusion}

Large mobile networks can profit from energy savings. This can be
achieved by computing schedulers which turn off or on cells while
maintaining some optimality criteria.
In this work we have model a given portion of the mobile network as a
stochastic game, applied different methodologies and finally used the
tool \stratego to synthesize near optimal strategies which minimize
energy consumption while maintaining coverage.

We have implemented a simulator which replays historical data. We have
performed simulations for two large geographical areas in Aalborg,
Denmark.  Our initial results are encouraging, showing energy savings
from up to 10\% and showing the scalability of our approach.

\paragraph{Future Work}

Currently we have distributed controllers which do not communicate
with each other. It would be interesting to study cooperative
distributed controllers in this contexts.  Our traffic demand model is
quite abstract and could be refined if more is available. In
particular a forecasting model for the traffic development could be of
interest. In the same manner our optimization function is quite simple,
one could consider to optimize different KPI's.


\bibliographystyle{abbrv}
\bibliography{main}

\end{document}